# A singular Lambert-W Schrödinger potential exactly solvable in terms of the confluent hypergeometric functions

A.M. Ishkhanyan[1,2,3]

[1]Institute for Physical Research, NAS of Armenia, Ashtarak 0203, Armenia
[2]Armenian State Pedagogical University, Yerevan 0010, Armenia
[3]Institute of Physics and Technology, National Research Tomsk Polytechnic University, Tomsk 634050, Russia

We introduce two potentials explicitly given by the Lambert-W function for which the exact solution of the one-dimensional stationary Schrödinger equation is written through the first derivative of a double-confluent Heun function. One of these potentials is a singular potential that behaves as the inverse square root in the vicinity of the origin and vanishes exponentially at the infinity. The exact solution of the Schrödinger equation for this potential is given through fundamental solutions each of which presents an irreducible linear combination of two confluent hypergeometric functions. Since the potential is effectively a short-range one it supports only a finite number of bound states.

## 1. Introduction

The most known solutions of the Schrödinger equation are the solutions in terms of the hypergeometric functions. Among these, distinguished are the potentials for which all the involved parameters can be varied independently. The list of the currently known such potentials involves three potentials that are solved in terms of the ordinary hypergeometric functions - the Eckart [1], the Pöschl-Teller [2], and the third hypergeometric potential [3], and five potentials that are solved in terms of the confluent hypergeometric functions - the harmonic oscillator [4], the Coulomb [4-5], the Morse [6], the inverse square root [7], and the Lambert-W step [8] potentials.

Less discussed are the potentials admitting the solution in terms of the functions of the Heun class [9-16]. It has recently been shown that if one considers the potentials that are proportional to an energy-independent parameter and if the potential shape is independent of both energy and that parameter there exist in total 29 independent Heun potentials [10-12]. There are eleven independent potentials that admit the solution in terms of the general Heun functions [10], for nine independent seven-parametric potentials the solution is given in terms of the confluent Heun functions [11], there are three independent double-confluent and five independent bi-confluent Heun potentials [12] (the six-parametric Lamieux-Bose potentials [9]), and one tri-confluent Heun potential (the general five-parametric quartic oscillator) [12].

There is another interesting class of special functions - the class of functions that involve the derivatives of the Heun functions. These functions in general do not belong to the Heun class of functions [17-21]. Nevertheless, they can be treated as a part of the theory of the Heun functions because they have a representation involving a derivative of a (general or confluent) Heun function. These functions are the solutions of the *deformed* Heun equations which are the Heun equations with an additional *apparent* singularity [19-21]. It has been shown that by an anti-quantization procedure the deformed Heun equations generate the complete totality of the six Painlevé equations [20-22]. The parameters involved in the Painlevé equations [23-24] are invariants of the corresponding Heun equations. Consequently s-homotopic transformations do not affect the forms of the Painlevé equations. The transpositions of the singularities in the Heun equations modify the forms of the Painlevé equations; however, they conserve the Painlevé property (the only movable singularities are poles) [23-24] (see also §5.1 in [15]).

We have recently discussed the reduction of the Schrödinger equation to the deformed tri-confluent [7] and the deformed bi-confluent Heun [8] equations. As a result, we have constructed the solution for the the inverse square root [7] and the Lambert-W step [8] potentials in terms of the confluent hypergeometric functions. In this letter we consider the reduction of the one-dimensional Schrödinger equation to the deformed *double-confluent* Heun equation. In this way we derive two potentials admitting the solution of the one-dimensional Schrödinger equation in terms of the first derivative of the double-confluent Heun functions. Both potentials are explicitly written through the Lambert function $W(x)$ which is an elementary function implicitly defined through the equation $We^W = x$ [25-26].

One of the derived potentials presents a generalisation of the Lambert-W step-potential [8]. It turns out that the Schrödinger equation for this potential is exactly solvable in terms of the confluent hypergeometric functions because the involved double-confluent Heun function in this case is expressed in terms of the latter functions. The result is that the general solution of the Schrödinger equation is written through fundamental solutions each of which presents an irreducible linear combination of two confluent hypergeometric functions. A sub-potential of this potential is a singular potential that behaves as the inverse square root in the vicinity of the origin and vanishes exponentially at the infinity. Discussing the bound states supported by this singular potential, we derive the exact equation for the energy spectrum and show that in contrast to the Coulomb or the inverse square root potentials it supports only a finite number of bound states.

## 2. The potentials

Thus, we consider the reduction of the one-dimensional Schrödinger equation for a particle of mass $m$ and energy $E$ in the field of a potential $V(x)$:

$$\frac{d^2\psi}{dx^2} + \frac{2m}{\hbar^2}(E - V(x))\psi = 0, \qquad (1)$$

to a *deformed* double-confluent Heun equation. The double-confluent Heun equation is one of the four confluent reductions of the (general) Heun equation achieved via coalescence of its singularities [14-15]. The double-confluent Heun equation has two irregular singularities of rank 1 conventionally located at $z = 0$ and $z = \infty$. Though the equation has four irreducible parameters [14-15], however, for convenience we consider a form of this equation which differs from that applied in the standard references in that it is written in an auxiliary five-parametric representation. The double-confluent Heun equation that we start with is

$$u_{zz} + \left(\frac{\gamma}{z^2} + \frac{\delta}{z} + \varepsilon\right)u_z + \frac{\alpha z - q}{z^2}u = 0. \qquad (2)$$

If $\alpha \neq 0$ the equation obeyed by the following function involving the first derivative of the solution $u(z)$ of this equation:

$$w(z) = z^\delta e^{\varepsilon z - \gamma/z}\frac{du}{dz}, \qquad (3)$$

has an additional regular *apparent* singularity located at the finite point $z_0 = q/\alpha$:

$$\frac{d^2 w}{dz^2} - \left(\frac{\gamma}{z^2} + \frac{\delta - 2}{z} + \varepsilon + \frac{1}{z - z_0}\right)\frac{dw}{dz} + \frac{\alpha(z - z_0)}{z^2}w = 0. \qquad (4)$$

This is the form of the deformed double-confluent Heun equation the reduction to which of the Schrödinger equation (1) we consider. The technique for this reduction is based on the results of [12] (see also [11]) and follows the particular steps first developed for the quantum two-state problem in [27-29] and further applied to the Schrödinger equation in [7-8,10-11] and to the Klein-Gordon equation in [30]. The derivation lines are as follows.

The transformation of the dependent variable $\psi = \varphi(z)\,w(z)$ for a coordinate transformation $z = z(x)$ and a pre-factor $\varphi(z)$ given as

$$\varphi(z) = \rho(z)^{-1/2} \exp\left(\frac{1}{2}\int f(z)dz\right), \qquad (5)$$

where $\rho(z) = dz/dx$, reduces the solution of equation (1) to solving the equation [12]

$$I(z) = g - \frac{f_z}{2} - \frac{f^2}{4} = -\frac{1}{2}\left(\frac{\rho_z}{\rho}\right)_z - \frac{1}{4}\left(\frac{\rho_z}{\rho}\right)^2 + \frac{2m}{\hbar^2}\frac{E - V(z)}{\rho^2}, \qquad (6)$$

where $f(z)$ and $g(z)$ are the coefficients of the deformed double-confluent Heun equation (4) and $I(z)$ is the *invariant* [15] of that equation if it is rewritten in the *normal form* [31].

The basic assertion of [12] is that if a potential is proportional to an energy-independent parameter and has a shape which is independent of the energy and that parameter, then the logarithmic $z$-derivative $\rho'(z)/\rho(z)$ cannot have poles other than the finite singularities of the target equation to which the Schrödinger equation is reduced (see theorem (27) of [12]). It then follows that because the deformed double-confluent Heun equation (4) possesses two finite singularities, $z=0$ and $z=z_0$, the appropriate coordinate transformation $z(x)$ is of the form

$$\rho(z) = z^{m_1}(z-z_0)^{m_2}/\sigma \qquad (7)$$

with integer or half-integer $m_{1,2}$ and arbitrary scaling constant $\sigma$. With this, the pre-factor $\varphi(z)$ becomes (see equation (5))

$$\varphi(z) = z^{-\frac{m_1+\delta-2}{2}}(z-z_0)^{-\frac{m_2+1}{2}}e^{-\left(\varepsilon z - \frac{\gamma}{z}\right)/2}. \qquad (8)$$

To do the next step, we note that the invariant $I(z)$ of equation (4) is a sixth-degree polynomial in $z$ divided by $z^4(z-z_0)^2$. Now, following the recipe of [12], we apply the simultaneous limit $E,V \to 0$ in equation (6) and further match the first two terms on the right-hand side, that depend only on $\rho(z)$ (which is supposed both $E$- and $V$-independent), with the invariant in the vicinity of the singular point $z_0$:

$$I(z)\bigg|_{\substack{E,V\to 0 \\ z\to z_0}} \sim -\frac{1}{2}\left(\frac{\rho_z}{\rho}\right)_z - \frac{1}{4}\left(\frac{\rho_z}{\rho}\right)^2 \bigg|_{z\to z_0}. \qquad (9)$$

As a result, we arrive at a simple equation:

$$-\frac{3}{4} = \frac{m_2}{2} - \frac{m_2^2}{4}, \qquad (10)$$

from which we get that $m_2 = -1$ (it is checked that the other root of this equation $m_2 = 3$ leads to inconsistencies). Further, demanding the energy term $2mE/(\hbar^2\rho^2)$ multiplied by $z^4(z-z_0)^2$ be a sixth degree polynomial as the form of the invariant requires, we get $m_1 = 1, 3/2$ or $2$.

Finally, in order to match the potential term with the rest part of equation (6), we make the substitution

$$V(z) = \rho(z)^2 \frac{v_0 + v_1 z + v_2 z^2 + v_3 z^3 + v_4 z^4 + v_5 z^5 + v_6 z^6}{z^4 (z - z_0)^2}, \tag{11}$$

where $v_i$, $i = 0,1,2,3,4,5,6$ are constants. With this $V(z)$ and $\rho(z)$ given by equation (7), collecting the coefficients at powers of $z$ in the numerator of equation (6) and demanding the constants $v_i$ to be independent of each other we get that the choice $m_1 = 3/2$ causes inconsistencies while the other two choices, $m_1 = 1$ and $m_1 = 2$, produce two potentials.

If $m_{1,2} = (2,-1)$, the result reads

$$V(z) = V_0 + \frac{V_1 z^2 / z_0^2}{1 - z/z_0}, \quad z = -\frac{z_0}{W\left(-z_0 e^{(x_0 - x)/\sigma}\right)}, \tag{12}$$

where $W$ is the Lambert function and the parameters of the double-confluent Heun function involved in the solution of the problem are given as

$$(\gamma, \delta, \varepsilon, \alpha, q) = \left(-\delta z_0, \sqrt{\frac{8m\sigma^2 (-E + V_0)}{\hbar^2}}, 0, \frac{2m\sigma^2 V_1 / z_0}{\hbar^2}, \alpha z_0\right). \tag{13}$$

For $m_{1,2} = (1,-1)$ we have

$$V(z) = V_0 + \frac{V_1}{1 - z/z_0}, \quad z = -z_0 W\left(-e^{(x_0 - x)/(\sigma z_0)} / z_0\right) \tag{14}$$

with the corresponding parameters

$$(\gamma, \delta, \varepsilon, \alpha, q) = \left(0, 1 - \varepsilon z_0, \sqrt{\frac{8m\sigma^2 (-E + V_0)}{\hbar^2}}, \frac{2m\sigma^2 z_0 V_1}{\hbar^2}, \alpha z_0\right). \tag{15}$$

## 3. The solution for a singular Lambert-W potential

Examining now the final solution of the Schrödinger equation we note that for the first potential the involved double-confluent Heun function in general is not reduced to simpler special functions. One may examine the possibilities for quasi-exactly solvable reductions for specific choices of the involved parameters including the parameter $z_0$ that indicates the location of the additional apparent singularity of the deformed Heun equation (4). Presumably, such quasi-polynomial solutions can be constructed by termination of the (asymptotic) power-series expansions of the involved double-confluent Heun functions [14]. A further progress may be expected if one tries more advanced expansions in terms of the Kummer confluent hypergeometric functions (see, e.g., [32-34]). However, this point is out of the scope of the present treatment.

As regards the second potential, it presents a generalization of the exactly solvable Lambert-W potential [8]. Indeed, at $z_0 = -1$ we recover the Lambert-W *step-potential* [8]. A different shape is obtained by putting $z_0 = +1$:

$$V = V_0 + \frac{V_1}{1-z}, \quad z = -W\left(-e^{-(x-x_0)/\sigma}\right). \tag{16}$$

For $x_0 = -\sigma$ with a positive $\sigma > 0$ this is a *singular* potential defined on the positive half-axis $x > 0$. We note that this potential is a member of the Lambert-W family of the single-confluent Heun potentials [11-12]. In order the potential to vanish at infinity we put $V_1 = -V_0$ (figure 1). The potential is then finally written as

$$V = \frac{V_0}{1 + 1/W(-e^{-(x+\sigma)/\sigma})}. \tag{17}$$

The behavior of this potential in the vicinity of the origin resembles that of the inverse square root potential [7]:

$$V\big|_{x\to 0} = -\frac{\sqrt{\sigma/2}\,V_0}{\sqrt{x}} + O(1). \tag{18}$$

However, the potential is a short-range one because it vanishes at infinity exponentially:

$$V\big|_{x\to +\infty} \sim -V_0 e^{-\frac{x+\sigma}{\sigma}}. \tag{19}$$

Importantly, the integral of the function $xV(x)$ over the semi-axis $x \in (0, +\infty)$ is finite, hence, according to the general criterion [35-37] the potential supports only a finite number of bound states.

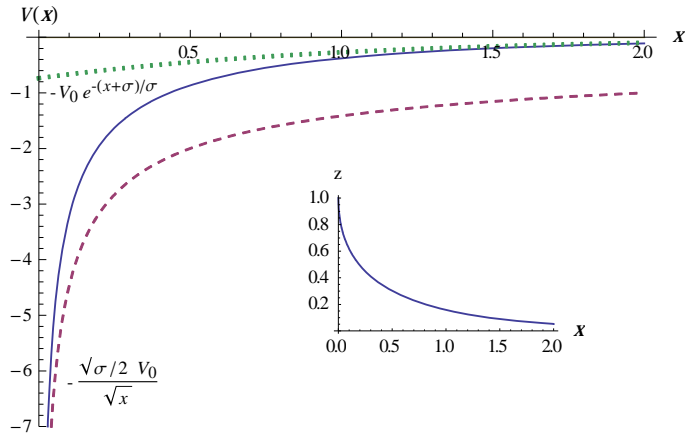

Fig.1. Potential (17) for $V_0 = 2$, $\sigma = 1$, $m = \hbar = 1$ (solid line). The dashed line presents the asymptote (18) for $x \to 0$ and the dotted line stands for the exponential asymptote (19) for $x \to \infty$. The inset presents the coordinate transformation $z(x)$.

A useful feature of the solution of the Schrödinger equation for the potential (17) is that $\gamma$ in this case is zero. The involved double-confluent Heun function is then expressed in terms of the confluent hypergeometric functions. The resultant general solution of the Schrödinger equation for arbitrary (real or complex) parameters $V_0$ and $x_0, \sigma$ is written as

$$\psi(x) = z^{c/2} e^{-cz/2} \frac{du}{dz}, \qquad (20)$$

with
$$u = e^{z(c-s_0)/2} \left( C_1 \cdot {}_1F_1(a;c;s_0 z) + C_2 \cdot U(a;c;s_0 z) \right), \qquad (21)$$

where $z = -W\left(-e^{-(x-x_0)/\sigma}\right)$, $C_{1,2}$ are arbitrary constants, ${}_1F_1$ and $U$ are the Kummer and the Tricomi confluent hypergeometric functions, and the involved parameters are given as

$$a = -\frac{(c-s_0)^2}{4 s_0}, \quad c = \pm\sqrt{\frac{-8m\sigma^2 E}{\hbar^2}}, \quad s_0 = \pm\sqrt{\frac{8m\sigma^2(-E+V_0)}{\hbar^2}}. \qquad (22)$$

The solution applies for $E \neq 0$ (since $c=0$ if $E=0$). The zero-energy solution should be written in a different way - see the next section. We note that here any combination of signs for $c$ and $s_0$ is applicable. The different combinations of the signs provide different pairs of independent fundamental solutions. This feature may be useful in applications since depending on a specific physical context one may need a particular asymptotic behavior that in some cases can be achieved by choosing an appropriate combination of the signs.

A notable future of the derived solution is that each of the two fundamental solutions that compose the general solution is written as an irreducible linear combination of two confluent hypergeometric functions. Hence, this solution cannot be derived by applying to the Kummer confluent hypergeometric equation the ansatz $\psi = \varphi(z) u(z)$ with only one confluent hypergeometric function $u(z)$.

**4. Bound states**

Discussing the bound states supported by the potential (17), we note that because the potential exponentially vanishes at infinity, in contrast to the Coulomb [4] or the inverse square root [7] potentials, it supports only a finite number of bound states.

This is already guessed from the graphical representation of the exact spectrum equation shown in figure 2. This equation is derived by demanding the vanishing of the wave function both in the origin and at the infinity. We then get that if the plus signs for $c$ and $s_0$ are chosen in equations (22), then the bound state wave functions are given by equations (20),(21) with $C_2 = 0$. Explicitly, these wave-functions (not normalized) are written as

$$\psi_B = C_1 z^{c/2} e^{-s_0 z/2} \left( \frac{c-s_0}{2} {}_1F_1(a;c;s_0 z) + \frac{a s_0}{c} {}_1F_1(a+1;c+1;s_0 z) \right) \quad (23)$$

with $z = -W\left(-e^{-(x-x_0)/\sigma}\right)$, the energy eigenvalues being determined from the equation

$$F(E) \equiv 1 + \frac{s_0 - c}{2c} \frac{{}_1F_1(1+a;1+c;s_0)}{{}_1F_1(a;c;s_0)} = 0. \quad (24)$$

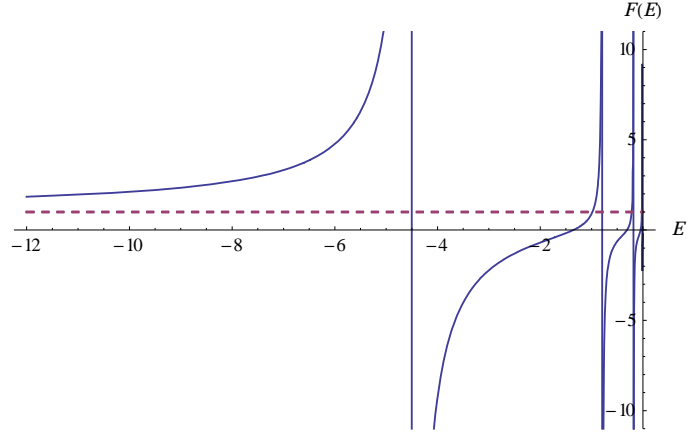

Fig.2. Graphical representation of the exact spectrum equation (24) for $V_0 = \sigma = 3$, $m = \hbar = 1$.

We note that for large negative energies $c, s_0 \to +\infty$ and $s_0 \to c$, hence, holds $F|_{E \to -\infty} = 1$. With this limit (shown in figure 1 by dashed line), it is understood that the number of the bound states is indeed finite.

The strict conclusion about the finiteness of the number of the bound states is achieved by checking the integral

$$I_B = \int_0^\infty r \left| V(x \to r\hbar/\sqrt{2m}) \right| dr. \quad (25)$$

It is a fundamental property of any spherically symmetric potential $V(r)$ for which this integral exists that there are only a finite number of bound states. Bargmann's inequality [35] gives the upper limit estimate for the number of the bound states as $n_l \leq I_B/(2l+1)$ for a given angular momentum $l$. Considering then the Schrödinger equation (1) as the $s$-wave reduction ($l=0$) for the spherically symmetric version of the potential (17), we have

$$n \leq I_B = \frac{m\sigma^2}{\hbar^2} V_0. \quad (26)$$

For the parameters $V_0 = \sigma = 3$, $m = \hbar = 1$ applied in figure 2, the estimate gives $n \leq I_B = 27$. Though Bargmann's inequality is the "best possible estimate" in the sense that it is possible to construct a potential (a chain of $n_l$ delta-potentials [35,36]) that saturates the limit, more strictly, "for a given $l$ potentials may be constructed which have a prescribed number $n_l$ of bound states for that angular momentum and for which $I_B$ approaches $(2l+1)n_l$ arbitrarily closely" [35], this does not imply that this bound provides a stringent limitation for all potentials [37-39]. Indeed, we see that for the particular potential we treat the inequality gives a rather overestimated result (figure 2 indicates that actually $n=3$). Other upper limits for the number of bound states have been found for particular potentials which give estimates more precise than those obtained from Bargmann's inequality [37-39].

Bargmann's estimate is considerably improved using Calogero's bound which is specialized for everywhere monotonically non-decreasing attractive central potentials [37]. This estimate implies that for strongly attractive potentials the number of bound states for a given angular momentum increases as the square root of the strength of the potential. For the potential we present the number of the bound states according to Calogero's estimate is

$$n \leq I_c = \sqrt{\frac{2m\sigma^2 V_0}{\hbar^2}}. \tag{27}$$

For the parameters applied in figure 2 this estimate is $n \leq I_c \approx 7.348$. It is seen that being more stringent, still, it overestimates the number of bound states by approximately a factor 2. The asymptotic result by Chadan for the limit of large coupling [38] provides a further tuning for the number of bound states equal to the half of that by Calogero, that is $n \leq I_c/2 \approx 3.674$.

We would like to conclude by noting that according to the general theory the exact number of bound states is equal to the number of zeros (not counting $r=0$) of the zero-energy solution which vanishes at the origin [35-39]. Since for $E=0$ the lower parameter $c$ of the involved Kummer confluent hypergeometric function vanishes, the solution (21) is not applicable for this case. The appropriate general solution of the Schrödinger equation for $E=0$ can be written as

$$\psi_{E=0} = \frac{d}{dz}\left( z e^{-\frac{s_0 z}{2}} \left( C_1 \cdot {}_1F_1(1+a;2;s_0 z) + C_2 U(1+a;2;s_0 z) \right) \right) \tag{28}$$

with $a = -s_0/4$, $s_0 = \sqrt{8m\sigma^2 V_0/\hbar^2}$ and $z = -W\left(-e^{-(x-x_0)/\sigma}\right)$. Choosing the constants $C_{1,2}$ so that this solution vanishes in the origin, we get the needed function for finding the exact number of bound states. It is then readily checked that for the parameters applied in figure 2 the number of nodes (with $r \neq 0$) of this function is three so that the number of bound states for these parameters is indeed three, in accordance with what is concluded from figure 2.

## 5. Discussion

Thus, we have introduced two potentials for which the exact solution of the one-dimensional Schrödinger equation is written in terms of the first derivative of a double-confluent Heun function. The derivation is achieved by reduction of the Schrödinger equation to a deformed double-confluent Heun equation that involves an additional apparent singularity as compared with the double-confluent Heun equation from which it originates. The potentials are members of a family of the single-confluent Heun potentials defined through the Lambert-W function which is an elementary function implicitly defined by an algebraic equation.

One of the potentials is a singular potential that behaves as the inverse square root in the vicinity of the origin and vanishes exponentially at the infinity. We have seen that the exact solution of the Schrödinger equation for this potential is eventually written through independent fundamental solutions each of which involves an irreducible linear combination of two confluent hypergeometric functions.

A peculiarity of the presented Lambert-W potential is that it vanishes at infinity sufficiently fast to support only a finite number of bound states. Since the early fifties of the 20th century, starting from the pioneering works of Jost and Pais in 1951 [40] and Bargmann in 1952 [35], the determination of upper and lower limits on the number of bound states of a given potential in the framework of both non-relativistic and relativistic quantum mechanics always engaged the attention of theoretical and mathematical physicists. For the singular Lambert-W potential that we have reported above the integral $\int_0^\infty rV(r)dr$ is finite so that it definitely supports only a finite number of bound states. The potential then suggests an immediate analytic test for further developments towards more elaborate estimates for the number of the bound states.

We would like to conclude by noting that the derived solution of the particular deformed double-confluent Heun equation may itself be of interest as an extension of the rather limited list of closed-form solutions of the equations of the Heun class in terms of the hypergeometric functions [41-42]. Because of the extremely wide appearances of these equations in contemporary mathematics, physics and engineering (see, e.g., [14-16] and references therein) such reductions may find applications in many branches of core research. Apart from the immediate usage in treating linear quantum problems such as the recently reported solutions of the Schrödinger equation for the inverse square root [7] and the Lambert-W step [8] potentials or the earlier discussions of the quantum two-state [43] and surface plasmon polariton [44] problems, among the possible applications it is worth to mention the direction towards treating specific nonlinear problems. For instance, owing to the known relationship between the (linear) Heun and the (nonlinear) Painlevé equations [20-22] one may expect that the derived solution can be applied to obtain an explicit solution for the corresponding non-linear Painlevé equation or at least to study the classical movement in the corresponding field as it had been done in [45]. Besides, different applications to other nonlinear quantum-mechanical problems can be envisaged, e.g., using the two-term variational ansatz involving a scaled solution to a resembling linear problem suggested in [46,47].

**Acknowledgments**

This research has been conducted within the scope of the International Associated Laboratory IRMAS (CNRS-France & SCS-Armenia). The work has been supported by the Armenian State Committee of Science (SCS Grants No. 13RB-052 and No. 15T-1C323) and the project "Leading Russian Research Universities" (Grant No. FTI_120_2014 of the Tomsk Polytechnic University).